\documentclass[aps,pra,twocolumn,eqsecnum,amsmath,superscriptaddress,showpacs]{revtex4}

\usepackage{graphicx}
\usepackage{longtable}
\usepackage{epsfig}
\usepackage{dcolumn}
\usepackage{bm}
\usepackage{amssymb}
\usepackage{multirow}
\usepackage{times,color}
\usepackage{hyperref}

\begin{document}

\title{Quantum phase transitions in spin-1 compass chains}

\author {Guang-Hua Liu}
\affiliation{Department of Physics, Tianjin Polytechnic University, Tianjin 300387, People's Republic of China}

\author {Long-Juan Kong}
\affiliation{Department of Physics, Tianjin Polytechnic University, Tianjin 300387, People's Republic of China}

\author {Wen-Long You}\email{wlyou@suda.edu.cn}
\affiliation{College of Physics, Optoelectronics and Energy, Soochow
University, Suzhou, Jiangsu 215006, People's Republic of China}

\begin{abstract}
The ground-state phase diagram and quantum phase transitions (QPTs) in a spin-1 compass chain
are investigated by the infinite time-evolving block decimation (iTEBD) method. Various phases are discerned by energy densities, spin correlations and entanglement entropy. A generalized string correlator is found to be capable of describing the nonlocal string order in the disordered phase. Furthermore, in the noncritical disordered phase, the spin-spin correlations are found to decay exponentially. Except for a multicritical point ($J_{1}=0$, $J_{2}=0$), the QPTs are determined to have second-order characters. In addition, the central charges on these critical phase boundaries are determined to be $c$=1/2, therefore these QPTs belong to the Ising universality class.


\end{abstract}

\maketitle

\section{Introduction}
\label{intro}

Compass models have been the subject of intensive theoretical and experimental studies \cite{Nussinov15}, fostered especially by the Kitaev physics \cite{Kitaev}. The active interest of compass Hamiltonians is motivated by
the essential role of orbital
degrees of freedom in transition-metal oxide and chiral degrees of freedom in frustrated magnets, as well as mathematical models for
topological quantum computing \cite{Levin05,Nayak08}. The distinctive property is that the couplings between the internal spin (or spin-like) components are inherently spatially dependent. Considering electrons in partially filled transition-metal $d$ shells that can
emerge in transition-metal compounds, so-called $e_g$ and $t_{2g}$ orbital degrees give rise to two-flavor compass models (for $e_g$) and to three-flavor compass models (for $t_{2g}$).
The famous Kitaev model is a special type of compass model, which is
defined on a honeycomb lattice. Intensive studies of Kitaev model actuate the progressive understanding to topological quantum
orders and topological excitations \cite{Feng07}.
It was natural to extend the direction-dependent interactions to bond-dependent couplings in one dimension. A spin-1/2 one-dimensional (1D) variants of compass model (sometimes also referred to as the 1D Kitaev model) was defined on a chain, in which nearest-neighbor interactions sequentially toggle between $S^{x}_{2i-1} S^{x}_{2i}$ and $S^{z}_{2i} S^{z}_{2i+1}$ variants for odd and even bonds. Two-spin correlation functions are found to be extremely short-ranged \cite{Baskaran07,You12}, indicating a quantum spin liquid state. However, the ground-state properties and the low-energy excitations of spin-$S$ ($>$1/2) compass models are elusive and deserve  an investigation.

One consensus in the study of quantum many-body systems is that the properties of spin models where $S$ is integer and half-odd integer are qualitatively different, especially after the Haldane's pioneering work \cite{Haldane1,Haldane2}. For decades the theorists knew that the classical N\'{e}el ground state is favored by Heisenberg antiferromagnetic (AF) chain with half integer spin $S$, while cannot play a similar role in that with integer spin $S$.
Numerous theoretical and numerical studies manifest that the ground state of the 1D $S$ = 1 Heisenberg antiferromagnet is separated from all excited states by a finite spin gap \cite{White93}, and thus
any spin-spin correlation is quenched in the Haldane phase of spin-1 Heisenberg model.
Instead, a hidden $Z_2$ $\times$ $Z_2$ symmetry breaking takes place \cite{Kennedy,Takada} and a nonlocal string order was discerned \cite{Nijs,Tasaki}.
The underlying physics of Haldane chains is fairly well
understood both in theory and experiments. For example, ¡°large-$D$¡± phase in $S$=1 AF Heisenberg chain with a strong planar anisotropy was realized in Ni(C$_2$H$_8$N$_2$)$_2$Ni(CN)$_4$ \cite{Verdaguer95} and
NiCl$_2$-4SC(NH$_2$)$_2$ \cite{Zvyagin07}. Using trapped ions to implement spin-1 XXZ AF chains as an experimental tool to explore the Haldane phase was proposed recently \cite{Cohen14}. In order to elucidate the properties of the spin-1 Heisenberg chain, the spin-1/2 ferromagnetic-antiferromagnetic (FA) alternating chain was proposed by Hida \cite{Hida,Hida2}, and a nonzero string order was also observed. It indicates that the string order parameter originally defined for the spin-1 Heisenberg chains can be generalized to the spin-1/2 FA alternating chain. Recently, a spin-1/2 Heisenberg-Ising (HI) alternating chain proposed by Lieb \emph{et al.} \cite{Lieb} was reinvestigated, and nonzero string orders were reported in the disordered phase \cite{Liu1}. However, relatively little is known about the magnetic properties and particularly the elementary excitation spectrum of non-Haldane $S$=1 AF chains. For instance, the $Z_2$$\times$$Z_2$ symmetry is fully broken in the Haldane phase, and hence the string order parameters are nonzero in both $x$- and $z$- directions. When the anisotropy is taken into account, the spin chains only have a $Z_2$ symmetry corresponding to the rotation of $\pi$ around a given axis \cite{Berg}. If the $Z_2$ symmetry in the ground state is broken, whether the string order parameter along the given axis becomes nonzero is not clear \cite{Hatsugai}. Therefore, it is also an interesting issue to explore the existence of the string correlators in spin-1 chains with lower symmetries than Heisenberg chain.

In the present paper, we would like to deal with a spin-1 compass chain, as a reminiscent of the much studied spin-1/2 counterpart \cite{Brzezicki07,Brzezicki09,You08,You10,You11,Eriksson,Subrahmanyam}.
The model describes the competing interactions between odd bonds and even bonds,
\begin{equation}
\hat{H} = \hat{H}_{\rm odd} + \hat{H}_{\rm even},
\label{Hamiltonian}
\end{equation}
in which
\begin{eqnarray}
\hat{H}_{\rm odd}&=&\sum^{N'}_{i=1}  J_1 S^{z}_{2i-1} S^{z}_{2i}+  J_2 S^{x}_{2i-1} S^{x}_{2i} , \label{Ham1} \\
\hat{H}_{\rm even}&=&\sum^{N'}_{i=1} L_1 S^{z}_{2i} S^{z}_{2i+1}.  \label{Ham2}
\end{eqnarray}
Here $S_{i}^{x}$ and $S_{i}^{z}$ are the $x$- and $z$- components of the spin-1 operators on the $i$th site. $N=2N'$ is the number of the sites. Parameters $J_1$, $J_2$, and $L_1$ denote the strength of nearest-neighbor spin couplings. In the following, we set $L_1$ to be unit. Spin operators in Hamiltonian (\ref{Hamiltonian}) obey the SU(2) algebra $[S_i^a, S_j^b]=i\delta_{ij} \epsilon_{abc}S_j^c$, for $a$,$b$,$c$=$x$,$y$,$z$ and $(\vec{S}_j)^2=S(S+1)=2$. However, Hamiltonian (\ref{Hamiltonian}) breaks the global spin-rotation symmetry, which is enjoyed by the conventional Heisenberg model. Specially, as the transverse coupling $J_2$ becomes zero, it is reduced to a spin-1 bond-alternating Ising chain. We note that a uniform $S$=1 Ising chain with single-ion anisotropy was solved rigorously \cite{Oitmaa,Yang1,Yang2}, and
a bond-alternating spin-1 Ising chain with single-ion anisotropy can also be solved exactly by means of a mapping to the spin-1/2 Ising chain and the Jordan-Wigner transformation \cite{Wu2}.
If the spin-1 operators in Hamiltonian (\ref{Hamiltonian}) are replaced by spin-1/2 ones, its ground state was found to be characterized by nonlocal string orders and doubly degenerate entanglement spectrum \cite{Motamedifar,Liu}. The double degeneracy of the entanglement spectrum reflects that the geometric bond space supports a projective representation of global symmetries, for instance, the bond centered inversion symmetry \cite{Pollmann-2010}, the rotational SO(3) symmetry \cite{Li-2013}.

The purpose of the present paper is twofold. First, we would like to obtain the ground-state phase diagram and discuss the quantum phase transitions (QPTs) in the 1D spin-1 model with both anisotropy and bond-alternation. Second, we want to examine the existence of the string orders in such a model. As will be shown below, four different ground-state phases, i.e., a ferromagnetic (FM) phase, a stripe (SP) phase, an AF phase, and a disordered phase,  will be distinguished by order parameters. Furthermore, the ground-state properties of the disordered phase will be discussed in detail. A generalized string correlator is found to be capable of describing the string orders in the disordered phase.

\section{Numerical Method}
\label{sec2}

Since the Jordan-Wigner transformation does not apply for spin-1 system any more, we will implement the infinite time evolution bond decimation (iTEBD) method \cite{Vidal} for the following investigation.
The iTEBD technique can accurately determine a variational wave function in the matrix product state (MPS) form.
The ground-state wavefunction $| \psi_g\rangle$ of Hamiltonian (\ref{Hamiltonian}) in period-two MPS form can be formally written as
\begin{equation}
| \psi \rangle = {\rm Tr} \left(\prod_{i=1}^{N'} \Gamma^{a} (m_{2i-1}) \Lambda^{a}
\Gamma^{b} (m_{2i}) \Lambda^{b}\right)|..., m_{2i-1}, m_{2i}, ...\rangle.
\label{wavefumction}
\end{equation}
Here $m_{i}$ is the local spin physical index, and the $\Gamma^{a}$ and $\Gamma^{b}$ are 3-index tensors on odd and even sites, respectively. $\Lambda^{a}$ ($\Lambda^{b}$) is a $\chi$ by $\chi$ ($\chi$ is the cut-off bond dimension) diagonal matrix of singular values on an odd (even) bond. To be concrete, an imaginary time evolution operator exp(-$\tau \hat{H}$) acts on an arbitrary initial state $| \psi_0\rangle$ to obtain the ground-state MPS wavefunction. In the limit $\tau \rightarrow \infty$, exp(-$\tau \hat{H} ) | \psi_0\rangle$ will converge to the ground state $| \psi_g\rangle$ of Hamiltonian $\hat{H}$.
When the $\delta\tau$ is infinitesimal, the evolution operator exp(-$\delta\tau \hat{H}$) can be expanded through a Suzuki-Trotter decomposition into a sequence of two-site gates $U^{[i,i+1]}$. The initial step $\delta \tau$ is set to be  $10^{-1}$, and then diminished to $10^{-8}$ gradually. In the practical imaginary time evolution process, two local tensors ($\Gamma^{a}$ and $\Gamma^{b}$) and two diagonal matrices ($\Lambda^{a}$ and $\Lambda^{b}$) should be updated repeatedly. The details of iTEBD method can be referred to original references \cite{Vidal}. Except for critical points, we find that $\chi$=20 can already provide rather accurate and converged results for this model. At the critical points, $\chi$=30 is adopted in our calculations. It should be pointed out that the criteria to select the value of $\chi$ is to assure a small cutoff error ($<10^{-9}$) in the  singular value decomposition process. In addition, the positions of pseudocritical points in the present model are dependent on $\chi$, so some scalings will be performed with $\chi$=10$\sim$50 to exactly determine both critical points.
Subsequently, based on the MPS ground state of Hamiltonian (\ref{Hamiltonian}), one can easily extract the expectation values of various quantities, such as energy densities and spin correlations. For instance, the ground-state energy per site is given by
\begin{eqnarray}
e_{i}= \frac{1}{N} \langle \psi_g |\hat{H} | \psi_g\rangle.
\end{eqnarray}

Meanwhile, it is especially convenient to excavate the entanglement using iTEBD algorithm. Quantum entanglement has been shown to be an efficient tool to describe QPTs in quantum systems \cite{Nielsen,Huang}. When the MPS is gauged
to its canonical form, one can cut any bond in the system, and obtain a Schmidt decomposition as
\begin{equation}
| \psi \rangle =\sum_{\alpha=1}^{\chi}\vert \phi_\alpha^{L} \rangle \Lambda_\alpha \vert \phi_\alpha^{R} \rangle,
\label{Schmidt_decomposition}
\end{equation}
where $\vert \phi_\alpha^{L} \rangle$ ($\vert \phi_\alpha^{R} \rangle$) represents the orthonormal bases of the
semi-infinite subsystem to the left (right) of the broken bond, and $\Lambda$ is a diagonal matrix. The diagonal elements 
of $\Lambda^2$ is called as entanglement spectrum \cite{Li08}. The bipartite entanglement of the semi-infinite chain ($S_{\rm B}$) can be directly read out from the entanglement spectra,
\begin{equation}
S_{\rm B} = - \sum_{\alpha=1}^{\chi} \Lambda_{\alpha}^{2} {\rm log}_2 \Lambda_{\alpha}^{2}.
\label{Definition Of Entanglement}
\end{equation}
Given the two-period MPS, two kinds of bipartite entanglement $S_{2i-1,2i}$ and $S_{2i,2i+1}$ are defined by $\Lambda^{a}$ and $\Lambda^{b}$ in Eq.(\ref{wavefumction}), corresponding to a cut on odd and even bonds, respectively. Besides, the block entanglement
is of great interest in many studies, given by
\begin{equation}
S_{\rm L}=-{\rm Tr} [\rho_{\rm L}{\rm log_{2}}\rho_{\rm L}],
\label{blockE}
\end{equation}
where $\rho_{\rm L}$ is the reduced density matrix of $L$ adjacent spins submerged at an infinite chain. $S_{\rm L}$ measures the bipartite entanglement between the center block with $L$ spins and the remaining environmental spins.

\section{Phase diagram and QPTs}
\label{sec3}

\begin{figure}
\includegraphics[width=8cm]{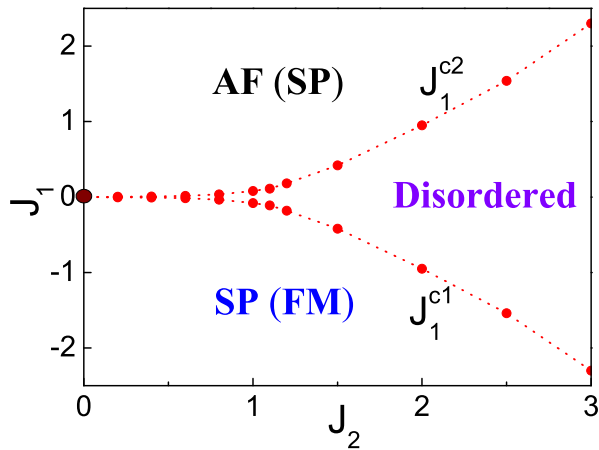}
\caption{(Colour online) Phase diagram of spin-1 compass chain. A SP (FM) phase, an AF (SP) phase and a disordered phase exist for $L_1$=1 ($L_1$=-1). }
\label{Fig1}       
\end{figure}

First, we would like to consider the case when the longitudinal Ising interactions on even bond is AF, i.e., $L_1$=1.
Specifically, Hamiltonian (\ref{Hamiltonian}) with $J_{2}$=0 is reduced to an Ising bond-alternating chain. Its ground
state becomes an ideal N\'{e}el state $\vert \cdots,1,-1,1,-1,\cdots \rangle$ as $J_{1}>0$, and the ground-state energy per site $e_{i}$=$-(J_1+L_1)/2$. When $J_{1}<0$, spins on odd bonds are parallel,
but spins on even bonds become antiparallel due to the AF couplings on even bonds. Therefore, an ideal stripe antiferromagnetic (SPAF)
state with configuration $\vert \cdots,1,1,-1,-1,\cdots \rangle$ is induced and $e_{i}$= $(J_1-L_1)/2$. One can see that $e_{i}$ undergoes a level crossing at $J_1=0$, indicating a first-order phase transition occurs. When the transverse interactions on odd bonds ($J_{2}$ terms) are switched on, the quantum fluctuations increase and an intermediate disordered phase is induced. The whole phase diagram is provided in Fig.~\ref{Fig1}, and three different phases, i.e., a SP phase, an AF phase, and a disordered phase, can be observed. Two second-order critical lines ($J_{1}^{c1}$ and $J_{1}^{c2}$) separate the disordered phase from the other phases.

\begin{figure}
\includegraphics[width=8cm]{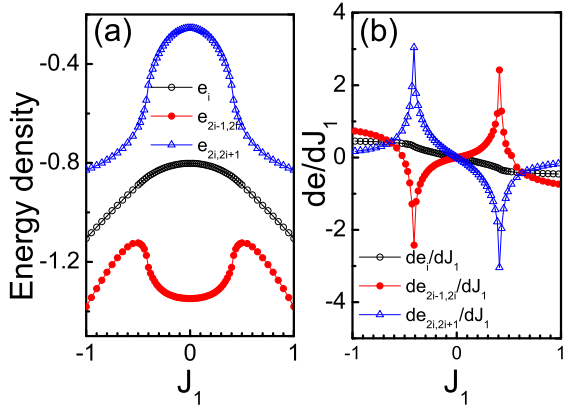}
\caption{(Colour online) Ground-state energy per site $e_{i}$ and ground-state energy per bond $e_{b}$ with $J_2=1.5$, $L_1=1$, $\chi$=20; (b) The first-order derivatives of $e_{i}$ and $e_{b}$.}
\label{Fig2}
\end{figure}

\begin{figure}
\includegraphics[width=8cm]{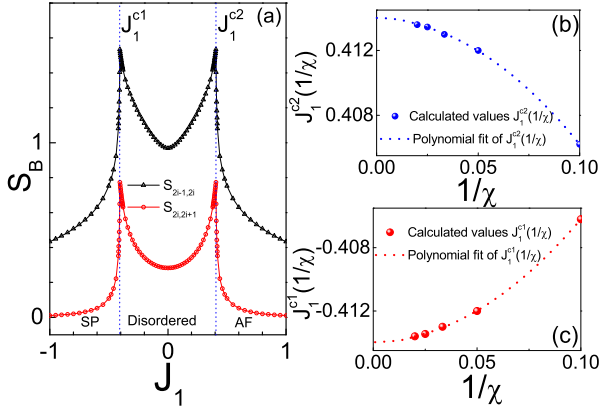}
\caption{(Colour online) (a) Bipartite entanglement on odd bond $S_{2i-1,2i}$ and even bond $S_{2i,2i+1}$ for $J_{2}=1.5$, $L_1=1$, $\chi$=20; Panels (b) and (c) show the scalings of critical points $J^{c2}_{1} (1/\chi)$ and $J^{c1}_{1} (1/\chi)$.}
\label{Fig3}
\end{figure}

In order to determine the phase boundaries, we can define the ground-state energy per bond including that on odd bonds  
 \begin{eqnarray}
  e_{2i-1,2i} = \frac{1}{N'}\langle \psi_g |\hat{H}_{\rm odd} | \psi_g\rangle
 \end{eqnarray}
and on even bonds 
 \begin{eqnarray}
 e_{2i,2i+1}=\frac{1}{N'} \langle \psi_g |\hat{H}_{\rm even} | \psi_g\rangle.
 \end{eqnarray}
We plot the above-mentioned energy densities as a function of $J_1$ for $J_2=1.5$ in Fig.~\ref{Fig2} (a). We find that $e_{i}$ changes smoothly with the increase of $J_1$, in contrast to the linear dependence and the associated level crossing for $J_2=0$.
Starting from very negative $J_1$, both $e_{2i-1,2i}$ and $e_{2i,2i+1}$ increase until $J_1$ approaches a critical value $J_{1}^{c1}$. $e_{2i-1,2i}$ continues to grow while $e_{2i,2i+1}$ begins to decline after exceeding $J_{1}^{c1}$. For $J_1>0$, another critical point locates at $J_{1}^{c2}$ $\equiv$ $-J_{1}^{c1}$ engraved by absolutely contrary characters with $J_{1}^{c1}$. More precisely, the curves of the ground-state energy are found to be symmetrical about $J_{1}$=0 [see Fig.~\ref{Fig2} (a)]. The symmetric spectrum implies the model (\ref{Hamiltonian}) is invariant with
respect to a gauge transformation to change the sign of $J_1$. This can be realized explicitly by a $\pi$-rotation of the spins at sites $4i-2$ and $4i-1$ around the $x$-axis.
Such criticality can be discriminated by taking the first-order derivative of bond energy [see Fig.~\ref{Fig2} (b)]. Although $d e_{i}/d J_{1}$ behaves continuously, both $d e_{2i-1,2i}/d J_{1}$ and $d e_{2i,2i+1}/d J_{1}$ exhibit two singularities with either minimum or maximum at $J_{1}^{c1}$ and $J_{1}^{c2}$.
One can speculate that
the first-order derivative of bond energy is in
fact a second-order derivative of site energy $e_{i}$. Take even bond
energy $e_{2i,2i+1}$ as an example, $d e_{2i,2i+1}/d J_{1}=d^2 e_{i}/d J_{1} d L_1$. Thus, these singular characters of $d e_{2i-1,2i}/d J_{1}$ and $d e_{2i,2i+1}/d J_{1}$ indicate both
QPTs should belong to the second-order category \cite{Liu}. Furthermore, the opposite trend of the energy density between odd and even bonds reveals a dimerized structure described by
$D$=$(e_{2i-1,2i}- e_{2i,2i+1})/e_{i}$ in the intermediate region \cite{Fath2001}.
In addition to the first-derivative of bond energy, the second-order pseudocritical points located at $J_{1}^{c1}$ and $J_{1}^{c2}$ can be seized by the divergent behavior of bipartite entanglement [see Fig.~\ref{Fig3} (a)], which provide another efficient way of identifying the exact critical points.
The scalings of $J_{1}^{c1}$ and $J_{1}^{c2}$ as a function of $1/\chi$ are provided in Figs.~\ref{Fig3} (b) and (c). Using polynomial fit $J_{1}^{c1}$ and $J_{1}^{c2}$ are determined to be $-0.414$ and $0.414$, respectively.

\begin{figure}
\includegraphics[width=8cm]{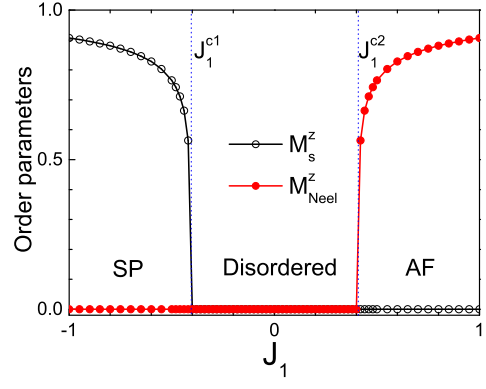}
\caption{(Colour online) Stripe AF order parameter
$M^{z}_{s}$ and N\'{e}el order parameter $M^{z}_{Neel}$ versus varying $J_{1}$ with $J_{2}=1.5$, $L_1=1$, $\chi$=20.}
\label{Fig4}
\end{figure}

\begin{figure}
\includegraphics[width=8cm]{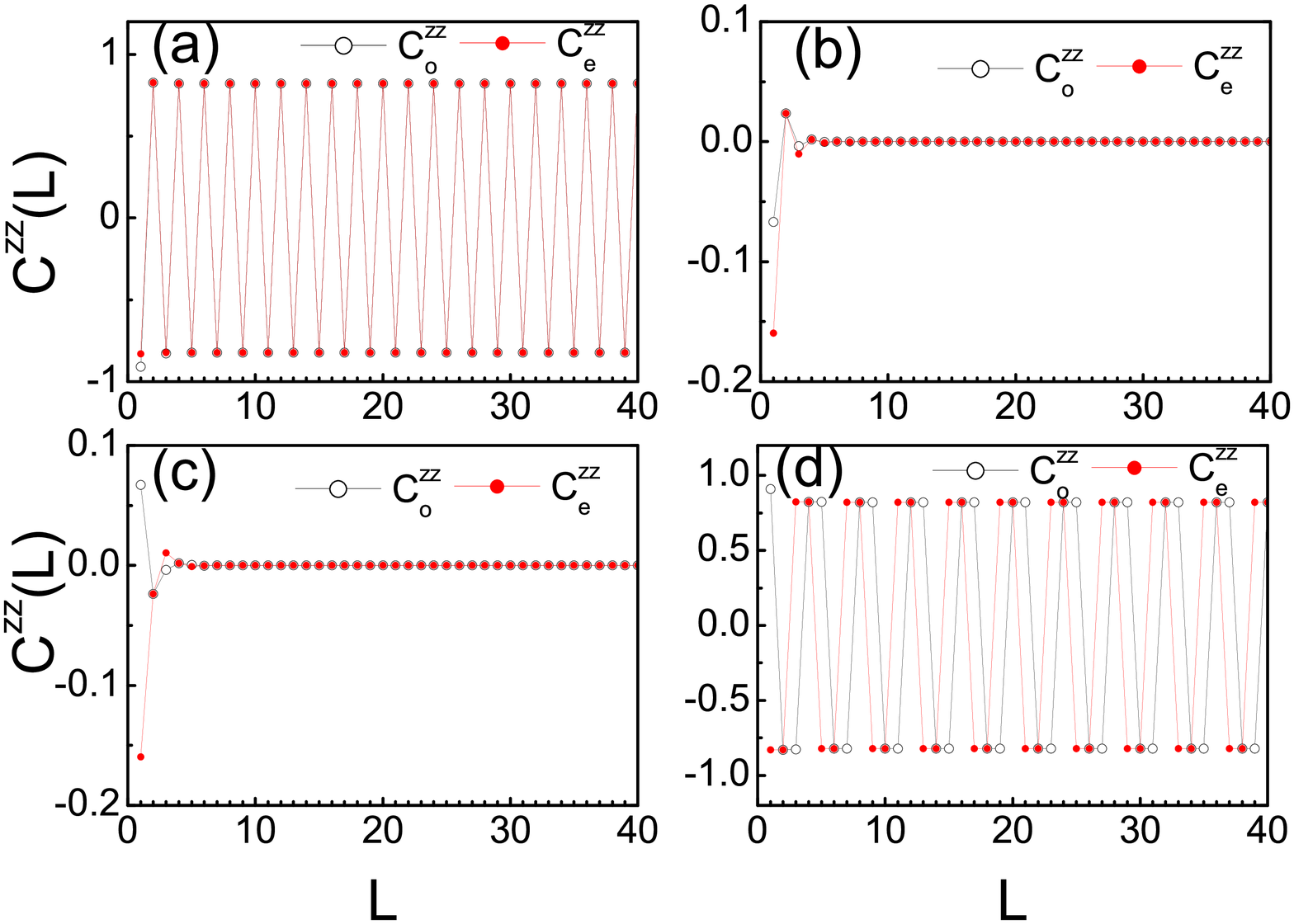}
\caption{(Colour online) Two-point correlation in dependence of separation $L$ for (a)$J_{1}$=1.0, $J_{2}$=1.5; (b) $J_{1}$=0.1 $J_{2}$=2.0; (c) $J_{1}$=-0.1,$J_{2}$=2.0; (d) $J_{1}$=-1.0, $J_{2}$=1.5. Parameters are as follows: $L_1=1$, $\chi$=20.}
\label{Fig5}
\end{figure}

Moreover, the local magnetizations $M^{\sigma}_{i}=\langle S^{\sigma}_{i}\rangle$ ($\sigma$=$x$, $y$, and $z$) versus varying $J_1$ are calculated (not shown), and they also hint the existence of three different phases separated by critical points $J_{1}^{c1}$ and $J_{1}^{c2}$. In the intermediate region $J_{1}^{c1}<J_{1}<J_{1}^{c2}$, the local magnetizations $M^{\sigma}_{i}$ ($\sigma$=$x$, $y$, and $z$) vanish completely. For $J_{1}<J_{1}^{c1}$ or $J_{1}>J_{1}^{c2}$, both transverse local magnetizations $M^{x}_{i}$ and $M^{y}_{i}$ remain null except $M^{z}_{i}$. A nonzero but unsaturated $|M^{z}_{i}|<1$ is observed. In order to discriminate these two phases with nonzero $M^{z}_{i}$, we calculate the SPAF order parameter $M^{z}_{s}$ = $\  |\sum_{i=1}^{N'} (-1)^{i}(M^z_{2i-1}+M^z_{2i})|/N$ and the N\'{e}el order parameter $M^{z}_{Neel}$ = $ \sum_{i=1}^{N} (-1)^{i} M^z_{i}/N$.  As is shown in Fig.~\ref{Fig4}, $M^{z}_{s}$ and $M^{z}_{Neel}$ become nonzero in the regions $J_{1}<J_{1}^{c1}$ and $J_{1}>J_{1}^{c2}$, respectively. It confirms the existence of a SP phase for $J_{1}<J_{1}^{c1}$ and an AF phase for $J_{1}>J_{1}^{c2}$. Besides, the long-range correlation  $C^{zz}(i,j)=\langle S^{z}_{i}S^{z}_{j}\rangle$ is calculated. The integer $i$ ($j$) specifies the position of the starting (ending) site, 
and accordingly $L$=$\vert j$-$i \vert$ measures the separation. 
$C^{zz}_{o}$ ($C^{zz}_{e}$) represents the two-point spin correlator of $z$ component for when starting site $i$ is odd (even). As shown in Figs.~\ref{Fig5} (a) and (d), distinctive nonzero and oscillating long-range correlations between two remote sites are observed in the SP and AF phases, but the oscillating patterns behave differently.
In the intermediate phase, we find that both $C^{zz}_{o}$ and $C^{zz}_{e}$ decay exponentially [see Figs.~\ref{Fig5} (b) and (c)], indicating such phase should be disordered, where only short-range spin correlations survive.

\begin{figure}
\includegraphics[width=8cm]{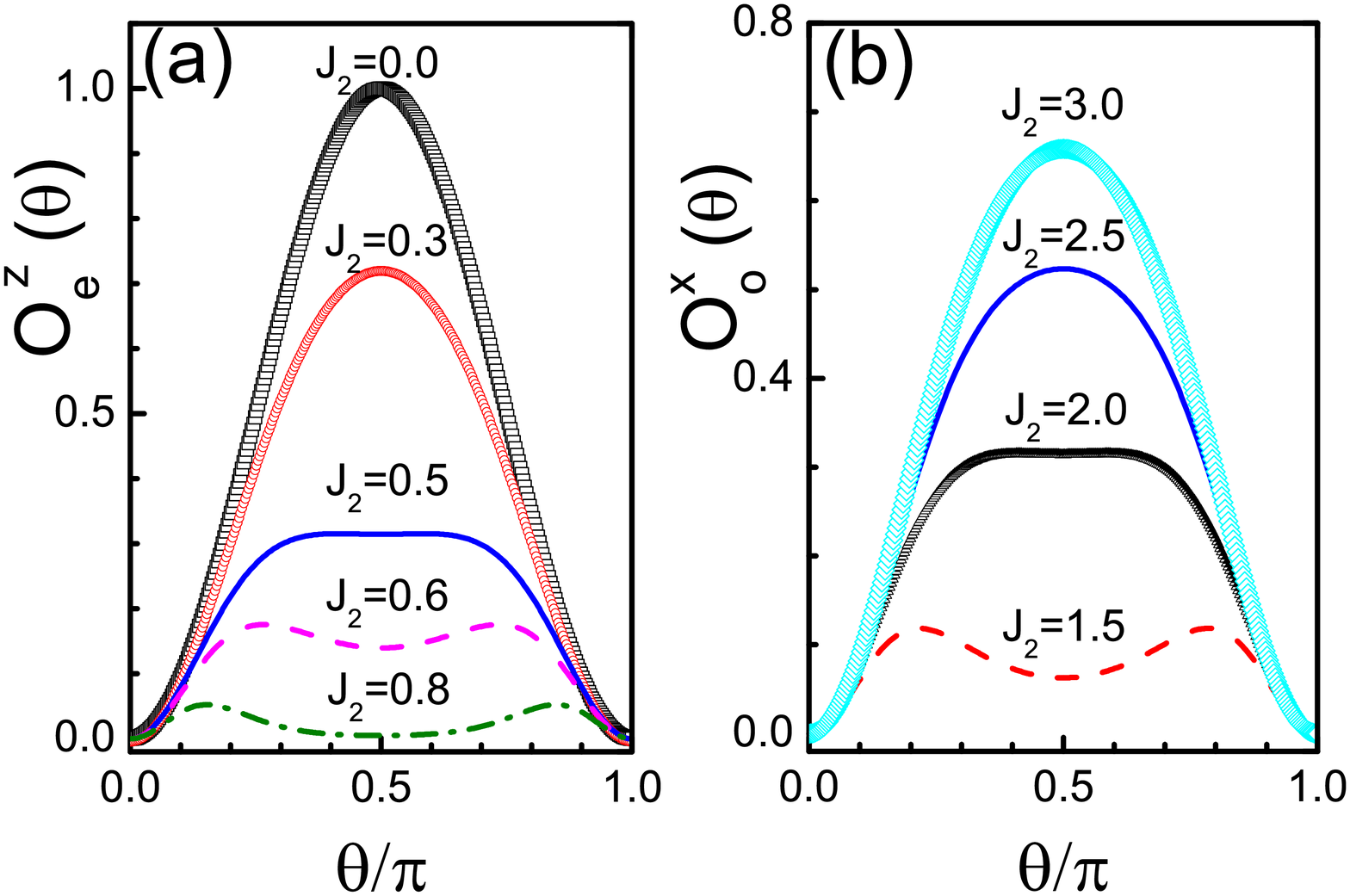}
\caption{(Colour online) Modified string correlators in the disordered phase versus varying $\theta$. (a) longitudinal even-string correlator ($O^{z}_{e}$) with $J_{2}\leq 1.0$; (b)transverse odd-string correlator ($O^{x}_{o}$) with $J_{2}\geq1.0$. Parameters are as follows: $J_{1}=0$, $L_1=1.0$, $\chi$=20.}
\label{Fig6}
\end{figure}

A natural question arises whether the disordered phase of spin-1 compass chain bears any resemblance to the Haldane phase of spin-1 Heisenberg chain, which is also disordered
and can be well characterized by a finite string order parameter \cite{Nijs,Tasaki}. The string correlators are initially defined as
\begin{equation}
O^{\sigma}_{s}=-{\rm lim}_{|i-j|\rightarrow\infty}\langle S^{\sigma}_{i} e^{i\pi\sum^{j-1}_{k=i+1} S^{\sigma}_{k}} S^{\sigma}_{j} \rangle,
\label{stringorder}
\end{equation}
where $\sigma$=$x$, $y$, and $z$. Afterward, a generalized version of string correlators was given by
\begin{equation}
O^{\sigma}_{s}(\theta)=-{\rm lim}_{|i-j|\rightarrow\infty}\langle S^{\sigma}_{i} e^{i\theta\sum^{j-1}_{k=i+1} S^{\sigma}_{k}} S^{\sigma}_{j} \rangle,
\label{modified string}
\end{equation}
where the special angle $\pi$ in the phase factor was replaced by a general angle $\theta$. Such extension was adopted by Tosuka \emph{et al.} to capture the hidden topological orders in the $S$=1 dimer phase \cite{Totsuka}, in which the maxima locate at $\theta$=$\pi/2$ and $3\pi/2$. We suspect that such a generalized string correlators can also provide us some clues about how to characterize the disordered phase of the present model. Considering the bond-alternation effects, we adopt two kinds of string correlators, i.e., the so-called odd- and even-string correlators. The odd-string correlators $O^{\sigma}_{o} (\theta)$ are defined by starting from a odd site $i$ to an even ending site $j$, and the even-string correlators $O^{\sigma}_{e} (\theta)$ are calculated from an even site $i$ to a odd site $j$. In the SP and AF phases, we find that $O^{x (y)}_{o}$ and $O^{x(y)}_{e}$ vanish absolutely. However, trivial nonzero string correlators $O^{z}_{o}$ and $O^{z}_{e}$ are observed in the SP and AF phases, whose maxima locate at both $\theta=0$ and $\theta=\pi$. Such trivial string orders are induced directly by the long-range longitudinal magnetic orders. In the disordered phase, we find that $O^{y}_{s}$ remains zero, while nontrivial string correlators $O^{z}_{e} (\theta)$ [$O^{x}_{o} (\theta)$] survive as a token of topological correlation for $J_{2}<L_1$ ($J_{2}>L_1$); cf. Figs.~\ref{Fig6} (a) and (b). The generalized string correlators are observed to be nonzero as $0<\theta<\pi$, but vanish absolutely at $\theta=0$ or $\pi$. Therefore, such nonlocal string orders can not be detected directly by the original definition in Eq. (\ref{stringorder}).
Specifically, as shown in Figs.~\ref{Fig6} (a) and (b), $O^{z}_{e} (\theta)$ and $O^{x}_{o} (\theta)$ become maximum at $\theta=\pi/2$ when $J_{2}$ is far away from 1.0 ($J_{2}$=0.0, 0.3, 2.5, and 3.0 are selected as examples), but become smaller and smaller as $J_{2}\rightarrow1.0$ ($J_{2}$=0.6, 0.8, and 1.5 are selected as examples). Note that $O^{z}_{e} (\theta)$ and $O^{x}_{o} (\theta)$ will not coexist in the disordered phase, since $Z_2$ symmetries along $z$ and $x$ axes are broken separately, which leads to the asynchronous emergence of the nonzero $O^{z}_{e}$ and $O^{x}_{o}$. Furthermore, if the string correlators are calculated from even to even or from odd to odd sites, they are always zero in the disordered phase.

\section{Discussions and Conclusions}
\label{sec4}

\begin{figure}
\includegraphics[width=8cm]{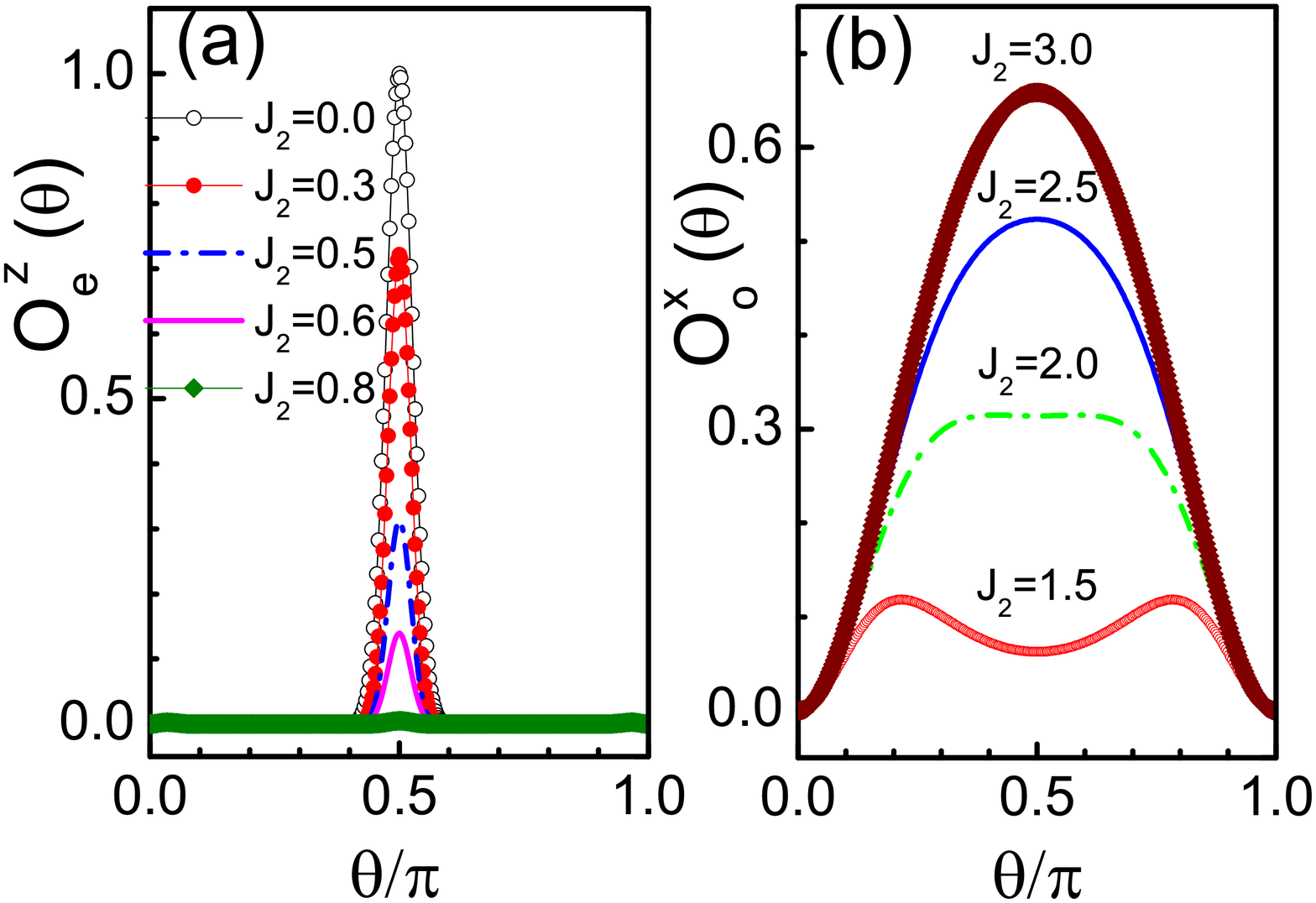}
\caption{(Colour online) Modified string correlators in the disordered phase versus varying $\theta$: (a) longitudinal even-string correlator ($O^{z}_{e}$) with $J_{2}\leq 1.0$; (b)transverse odd-string correlator ($O^{x}_{o}$) with $J_{2}\geq1.0$. Parameters are as follows: $J_{1}=0$, $L_1=-1.0$, $\chi$=20.}
\label{Fig7}
\end{figure}

By the iTEBD algorithm, the ground-state properties of spin-1 anisotropic and bond-alternating chains were investigated. Two second-order critical lines $J_{1}^{c1}$ and $J_{1}^{c2}$ have been recognized by the singularities of the second-derivative of energy density and the bipartite entanglement, and thus the ground-state phase diagram was determined. Nonzero generalized string correlators were observed in the disordered phase, where the long-range spin-spin correlations decay exponentially.
Our investigation focused on $L_1>0$, and we stress that the case
of FM Ising couplings on even bonds, i.e., $L_1<0$, is unitarily equivalent to the case with AF Ising coupling, i.e., $L_1>0$. To be concrete, Hamiltonian (\ref{Hamiltonian}) with
$L_1$=$-1$ and that with $L_1$=$1$ can be mutually transformed by a $\pi$-rotation of the spins at sites $4i-3$ and $4i-2$
around the $x$- or $y$-axis. As a consequence, one can easily obtain the ground-state phase diagram with $L_1$=$-1$. When all the spins at sites $4i-3$ and $4i-2$ are rotated over $x$-axis by $\pi$, the SP phase becomes a FM phase and the AF phase reduces to a SP phase. Therefore, the phase diagram with $L_1$=-1 is similar to that with $L_1$=1, and has the same phase boundaries (see the notations in the parentheses of Fig.~\ref{Fig1}). 
For the case with $L_1$=-1, we also calculate the $O^{z}_{e} (\theta)$ and $O^{x}_{o} (\theta)$ in the disordered phase in Fig.~\ref{Fig7}. It must be noted that  the string correlators in Fig.~\ref{Fig7} are directly calculated by the wavefunction with $L_1$=$-1$ without transformation. We find that string correlator $O^{z}_{e} (\theta)$ [see Fig.~\ref{Fig7} (a)] becomes narrower than the counterpart in Fig.~\ref{Fig6}. The long-range spin-spin correlations are also found to decay exponentially in the disordered phase for $L_1$=$-1$.

\begin{figure}
\includegraphics[width=7cm]{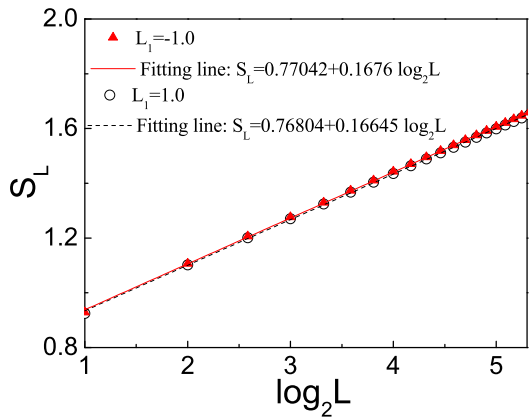}
\caption{(Colour online) Block entanglement $S_{L}$ at criticality for $L_1$=1 and $L_1$=-1. The black dashed and the red solid lines denote the corresponding fitting lines. Parameters are as follows: $J_{1}$=0.414, $J_{2}$=1.5, $\chi$=30.}
\label{Fig8}
\end{figure}

The QPTs from the disordered phase to the AF, SP, and FM phases have second-order characters. In order to uncover the further properties of these QPTs, we calculate the block entanglement $S_{L}$. Except for two critical lines, all the phases are found to be noncritical and gapped. The block entanglements in these phases are found to become saturated quickly as the block size $L$ increases, and well satisfy the entanglement area law \cite{Eisert}. Only on both critical lines ($J_{1}^{c1}$ and $J_{1}^{c2}$), the block entanglement $S_{L}$ violates the area law, and displays a logarithmic divergence behavior. As derived in Ref. \cite{Holzhey}, the divergent behavior can be described well by a logarithmic function
\begin{equation}
S_{L}=c_0+ \frac{c}{3} {\rm log}_{2} L,
\end{equation}
where $c_0$ is the intercept and $c$ denotes the central charge in the conformal field theory. Typical points ($J_{1}$=0.414, $J_{2}$=1.5, $L_1$=$\pm 1$) are selected from critical lines as examples. As is shown in Fig.~\ref{Fig8}, logarithmic divergences are observed, and the central charges are determined to be $c \simeq $1/2. It means the QPTs at both critical lines belong to the Ising universality class with central charge $c$=1/2, and can be described by a free fermionic theory \cite{Vidal2003}. This is quite similar to spin-1/2 compass models \cite{Liu}.

In conclusion, using the state-of-the-art iTEBD algorithm, we investigate the ground-state properties of the spin-1 compass chain. We give reliable evidences to depict the phase diagram and characterize each phase. Besides the ordered phase, the properties of the disordered phase were scrutinized, and the nonlocal string order are found to exist therein. However, string orders along $z$ and $x$ axes were found to exclude each other. Furthermore, the disordered phase is noncritical, and has exponentially decaying spin-spin correlations.

\begin{acknowledgements}
This work is supported by the Chinese National Science Foundation under Grant No. 11347008. W.-L.Y acknowledges support by the Natural Science Foundation of Jiangsu Province of China under Grant No. BK20141190 and the NSFC under Grant No. 11474211. All authors contributed equally to the present paper.
\end{acknowledgements}



\end{document}